# Three-dimensional nanoprinting via charged aerosol focusing

Wooik Jung[1,2†], Yoon-ho Jung[1,2†], Peter V. Pikhitsa[1], Jooyeon Shin[1,2], Kijoon Bang[1], Jicheng Feng[1] and Mansoo Choi[1,2*]

[1]*Global Frontier Center for Multiscale Energy Systems, Seoul National University, Seoul 08826, Korea*

[2]*Department of Mechanical and Aerospace Engineering, Seoul National University, Seoul 08826, Korea*

[†]*These authors contributed equally on this work*

*To whom correspondence should be addressed. E-mail: mchoi@snu.ac.kr*

Three-dimensional (3D) printing has attracted great attention due to the flexibility and practicality on 3D structuring for various applications. Many researches attempt to further scale down the 3D printing technique to take advantage of the unique optical, physical and chemical properties arising from 3D nanostructures for diverse applications including electronics (*1, 2*), energy device (*3, 4*), bioengineering (*5-9*) and sensing (*10*). Although various approaches have been developed such as DNA scaffold (*11, 12*), photo lithography (*13, 14*), electron-beam lithography (*15, 16*) and electrohydrodynamic (*17-20*) approaches, ensuring nanoscale resolution with high degree of freedom that is essential for further development towards the 3D nano-printing has been a challenge.

Previously, we developed a 3D nanoparticle assembly technique named Ion-Assisted Aerosol Lithography (IAAL) (*21-24*). IAAL is an aerosol-based manufacturing process, which guides charged nanoparticles following a distorted electric field induced by aerodynamic focusing lenses generated from accumulated ions on an insulating patterned layer. IAAL has great advantages of fabricating versatile 3D nanostructures for different applications (*25-27*).

Herein, we propose a novel 3D nano-printing concept for fabricating versatile 3D nanostructures that cannot be easily realized from existing methods. We apply a floating dielectric mask concept combined with 3D translation of piezoelectric nanostage to focus charged aerosols through convergent electrostatic field through apertures in the floating mask. Fine-tuning of the 3D translation speed and direction of the nanostage can determine the shape of the 3D nanostructures. Fundamental methodology and some of the results are presented in several conferences (*28-30*). Interestingly, we found two different regimes (3D growth mode and 3D writing mode) exist depending on the translation velocity of the stage. In the 3D growth mode, the shape of structure is determined by adjusting the stage translation speed according to the growth rate of the structure. Adding the horizontal movement of the stage enables the manufacturing of slanted structures in various angles. Precise 3-axis stage controlling can lead to complex 3D nanostructures. Relatively faster movement of stage would lead to 3D writing mode, in which nanoparticles cannot be accumulated upon the existing cluster and the movement of the stage makes a line of particles. This means that the multiple sweeping with a same orbit makes the writing of 3D nanostructures.

The general approach to produce a scaffold for the parallel structure growth in 3D is based on the floating dielectric mask with the array of holes (Fig. 1, A and B). The positively charged ions and aerosol nanoparticles created in a spark discharge (Fig. S2) are directed towards the mask and substrate by the electric field that controls the deposition process with the potential on the substrate. (Fig. S1 shows the experimental set-up.)

The role of the electric field is very complex: positive ions trapped by the mask out of the flow create a positively charged cushion on the mask. The cushion is shaped by the electric field from the ion distribution on the mask hole array so that the resulting electric field is repelling for the aerosol nanoparticles everywhere over the mask except the hole regions where it produces narrow funnels (Fig. 1A and Fig. S3). The nanoparticle stream focused within the funnels is directed to a given position on the substrate, controlled by the nanostage position. When the stage does not move, the pillars are growing in the direction of their tip (normal to the substrate) with the equal height controlled by the deposition time (Fig. 1, C to E). The mask hole array controls the resulting pillar array (Fig. 1F and Fig. S5). By alternating the type of nanoparticles during the deposition the pillars could be grown of complex material composition (see Fig. 1, G and H, for the pillar with upper half copper and lower half palladium.)

The full capacity of the 3D printing method is revealed when the stage moves controllably during the deposition process. The complete control over the structure growth shape becomes possible due to the further focusing of the electric field lines in the funnel onto exclusively the tip of the growing structure. The tip follows the electric field line and thus the latter plays the role of a drawing tool. One may recall Faraday's first drawings of the electric field lines with a pensile. Here we are drawing a 3D pensile with the electric field line.

Thus, a considerable simplification for designing various growing nanostructures in a 3D printing mode (see the simplest ones in Fig. 2B and Fig. S6) comes from the fact (supported by our numerical simulations (*31*) given in Fig. 2C) that the charged nanoparticles on average follow the electric field lines that pass through the mask hole (Fig. 2A) and end up on the tip of the growing structure. The tip, being the protruding element of the conductive surface of the agglomerate, collects coming nanoparticles because geometrically the concentration of electric field lines (normal to the agglomerate surface) increases near the tip (Fig. 2C). The guiding line is the one that passes through the central area of the mask hole, where nanoparticles are funneled by the local charge distribution of the focusing lens, and ends up on the tip (Fig. 2A).

Here lies an important conceptual difference between our approach and Diffusion Limited Aggregation (DLA) or Laplacian growth. In our approach the nanoparticles are forced to strictly follow the approximately vertical electric field lines due to focusing electrostatic lenses of the mask holes above the growing conductive surface of agglomerates. These lenses are controlled by the *fixed* charge distribution of ions trapped on the mask surface. Without these lenses the nanoparticles would perform the Brownian motion and stick to the growing agglomerate in the DLA manner, thus forming fractals. We indeed observe fractal-like growth when the focusing from the mask is not sufficient due to low concentration of deposited ions. It is because the only size restriction for stabilizing the DLA fractal growth is the size of the nanoparticles, while as we show, sufficient electrostatic focusing results in novel "persistence" length $L_p$. It is in the essence of our approach to put each electric field line (and thus the nanoparticle trajectory) to its place, determined by *both* the charge distribution on the silicon-nitride mask *and* the tip position. It is that lucky combination that turns the electric field line into a 3D writing tool.

The above-said leads to a simple growth description, in which the growth pattern in 3D printing can be completely controlled by a 2D stage motion protocol. As one can see from the schematic in Fig. 2, A and C, it is a fair approximation to select as the guiding streamline the one of the electric field lines that starts at the center of the hole in the mask (the center is chosen due to the symmetry, provided the mask hole is far enough from the growing structure and the substrate) and ends on the tip of the growing structure.

However, it is difficult to calculate general electric field configuration in 3D. Unlike 2D space where conformal properties of the complex plane give the possibility to use analytical methods and harmonic functions to calculate the potential and electric field lines (*32*), in 3D it is not generally possible. Still, for 3D printing one needs a sure control over the mask position and motion to draw a desired 3D structure with the electric field line. Below we show that phenomenology leads to a simple description of the field lines which is sufficient to control the 3D printing mode in creating rather complicated and counterintuitive structures (like the one in Fig. 2B). We believe that the stage being equipped with such a control can compete with the methods that use laser beam 3D writing, yet without the restriction on nonconductive materials and by moving the stage in 2D only, while the structure grows in 3D.

The geometry of the guiding electric field line in 3D can be calculated as follows. It is assumed that all electric field lines issue nearly normally from the mask surface (including the mask hole regions as far as the mask is sufficiently far from the growing structure) and bunch all together while focusing at the tip of the equipotential growing structure (Fig. 2C). The simplest phenomenological picture is obtained when solely the tip is considered as the point in 3D while neglecting the already growth structure and the substrate. Then it is easy to notice that the line bunching/focusing is governed by the electric field flux conservation equation $\pi R^2 \sigma = \frac{q}{4\pi} \int_0^\theta 2\pi \sin\theta d\theta = q \frac{1-\cos\theta}{2}$, where $R$ is the distance from the tip to the mask hole center along $x$ axis (Fig.2A); $\theta$ is the angle between the normal/vertical $z$ direction and the guiding line; $\sigma$ is the surface charge density on the mask, and $q$ is the effective "charge" on the equipotential tip surface that gives the phenomenological description of the line focusing on the tip. This leads to the equation

$$R = L_p \sin\frac{\theta}{2} \qquad (1)$$

where $L_p = \sqrt{\frac{q}{\pi\sigma}}$ is the phenomenological "persistence" length which determines how strong the focusing is. We picked out the term "persistence" length because of the close analogy with the elastic behavior while we are using the guiding electric field line as a flexible writing rod. In cases when the growth rate is nearly constant, one may also define the "response" time $= \frac{L_p}{v_g}$, where $v_g$ is the growth velocity of the structure.

Although looking quite innocent, Eq. (1) leads to surprisingly rich formulation. For application it is also important that the vertical distance between the mask and the stage plays little role provided the field lines issue nearly normally to the mask in the relevant region around each growing structure. That gives the possibility to make 3D structures by 2D motion of the stage.

Let us illustrate the phenomenology with two cases: a sudden 1D motion of the stage in $x$ direction with a constant stage velocity $v_s$ (different cases are given in Fig. 2B) and the sudden rotational 2D motion of the mask hole around $z$ axis at a constant angular frequency $\omega$ (a characteristic helix structure is shown in Fig. 3 along with its theoretical shape). The former leads to a steady-state growing structure inclined at some angle $\theta_s$, while the latter gives a steady-state helix which parameters (pitch angle $\theta_t$ and radius $R_t$) are tightly bound to the phenomenological persistence length $L_p$ and response time $\tau$.

For the stage motion along $x$ axis with velocity $v_s = v_g \sin \theta_s$, where $\theta_s$ is a given angle, the equations of tip growth motion in $x$ and $z$ directions are

$$\frac{dR}{dt} = v_g \sin \theta_s - v_g \sin \theta; \quad \frac{dz}{dt} = v_g \cos \theta. \tag{2}$$

Eq. (2) along with Eq. (1) describes typical relaxation behavior when angle $\theta = 0$ relaxes to its steady state value $\theta_s$ (Fig. 2, A and B) and the distance along the $x$ axis from the tip to the center of the mask hole approaches $L_p \sin \frac{\theta_s}{2}$ (so-called delay distance). One can estimate the persistence length $L_p \approx 1 \ \mu m$ by simply measuring the length of the "knee" region in Fig. 2B that manifests the transition from $\theta = 0$ to $\theta_s$. With the growth velocity $v_g \approx 50 \ \mu m/h$ one gets the estimate of $\tau \approx 1.2 \ min$.

Eq. (2) can be rewritten for the tip position $x_t$ in the general case of arbitrary stage position $x_s$.

It is interesting to note that counterintuitive downward growth (Fig. 2B) (when $\theta_s > \frac{\pi}{2}$) is always possible due to electric field line configurations at the initial condition when the mask hole is placed at the delay distance $L_p \sin \frac{\theta_s}{2}$ from the tip of the vertical pillar. In such situation the electric field line directs the tip growth downwards as is seen from Fig. 2A. At this steady-state, the stage velocity is $v_s = v_g \sin \theta_s$, being the same as for the upward growth steady-state value at angle $\pi - \theta_s < \pi/2$ because $\sin \theta_s \equiv \sin(\pi - \theta_s)$. However, now the delay distance at the steady-state downward growth $L_p \sin \frac{\theta_s}{2}$ is larger than the one for the upward growth $L_p \sin \frac{\pi - \theta_s}{2}$. As far as we verified this theoretical prediction experimentally (see Fig. 2B), this simple case of 3D printing brings confident in the theory in application to more sophisticated case of a helix. Furthermore, using movement of the stage, we can fabricate various structures. (Fig. S7 and S8)

A uniform field of the parallel helices is shown in Fig. 3, A and B (and Fig. S9). The case of a helix structure reveals the possibilities of our phenomenological theory to describe and predict the growth morphology in full 3D, given the stage motion in $x, y$ directions. Direct generalization on arbitrary stage motion is given in Supplemental Materials. The solution of general equations for a sudden steady state rotation of the mask hole with a constant angular frequency $\omega \approx 4\pi \ h^{-1}$ leads to the helix given in Fig. 3E (right panel) which shows a good correspondence to the experimental helix in Fig. 3E (left panel).

Now we only discuss a steady state helix which allows elementary treatment. From the simple geometry of the right-angle triangle in Fig. 3C (inset) three relations follow: $R = R_s \sin\varphi = L_p \sin\frac{\theta_t}{2}$, $R_t = R_s \cos\varphi$, $R_t \omega = v_g \sin\theta_t$, where $R_s$ is the radius of the rotation of the hole in the mask, $\varphi$ is the delay angle between the tip and the hole, $R$ as in Eq. (1) is the distance between the tip and the center of the mask hole.

After some simple algebra the relations give the solution for the helix radius $R_t$, pitch angle $\theta_t$, and the delay angle $\varphi$:

$$\frac{R_t}{R_s} = \frac{1}{\Delta w} \sin\theta_t \qquad (3)$$

where $\theta_t = 2\,\text{asin}\sqrt{g_{1,2}}$; $g_{1,2} = \frac{1}{2}\left(1 + \frac{w^2}{4} \mp \sqrt{\left(1+\frac{w^2}{4}\right)^2 - \Delta^2 w^2}\right)$; $w = \omega\tau$; $\Delta = R_s/L_p$, and the sign is chosen to be minus for $g_1$ if $\Delta < 1$ and plus for $g_2$ if $\Delta > 1$, so that the solution in Eq. (3) has two types of branches displayed in Fig. 3, C and D. The top view of the structure in Fig. 3E treated as a steady state structure obtained for $R_s = 1\ \mu m$, gives the ratio $\frac{R_t}{R_s} = 0.9$ from which, by using Eq. (3), the experimental growth velocity $v_g = 15.5\ \mu m/h$, and the experimental angular velocity $\omega \approx 4\pi\ h^{-1}$, one obtains $\tau = 0.072\ h = 4.3\ min$, persistence length $L_p = 1.12\ \mu m$, $\Delta = 0.8$, pitch angle $\theta_t = 41°$ (close to the pitch angle of $45°$ seen in Fig. 3E, tilt view), and the delay angle $\varphi = 25°$ (see Fig. 3E, top view, where the circle is incomplete up to the delay angle).

From Fig. 3C one can see that increase in dimensionless angular velocity decreases the helix radius and increases the pitch angle making the helix more compact. It is interesting to note that Eq. (3) at certain parameters contain the solutions that describe a stable steady state growth downwards (see Fig. 3D), similar to the downward growth from Fig.2B. For both types of branches in Fig. 3D there is the angular velocity above which the pitch angle overcomes the right angle and the helix may go downwards. Yet, there is a principal difference between the branches: the ones with $\Delta > 1$ stop existing above some angular frequency, which is frequently observed in experiment. However, even for this branch there always exists a narrow "window" when downward steady state growth is still possible (Fig. 3D).

As a universal drawing tool the electric field line is capable of switching to writing on a substrate. It happens when the horizontal stage velocity exceeds the growth rate in the vertical direction, normal to the substrate. (Fig. 4A and Fig. S10) Then it works as a brush tool to produce virtually any desirable shapes as any other brush including multiple passages over the same places to grow 3D structures in a different way (Fig. 4B). The material can be changed at any moment to produce composite structures (see Fig. 4, C, D and Fig. S11) of copper and palladium.

We used different slit-shaped holes of the mask to produce wall-like growth for the motionless mask instead of the tip-directed pillar growth. For the wall-like growth, considered as generalization of the tip-like growth, one can imagine that the point-like tip is spread along the line and instead of one leading electric field line we have a sheet. The various structures obtained (Fig. 5, A and B) basically follow the stencil hole shape, however, the strongly nonlinear nature of the growth stimulates wave-like structures along the wall line top especially pronounced for symmetrical stencil slit shapes (Fig. 5C). The phenomenological theory that we developed can be used to describe such shaped, yet it is out of the scope of the current paper.

The powerful and flexible method of 3D printing has been developed. The scaffold that it uses is the electric field configuration that has no restriction as to sizes, so one can imagine that it be scaled down to atomic sizes or up to mesoscopic ones. Broad material independence open the way for producing hybrid structures that are essential for electronic devices. The method contains three modes which are complementary: controlled *tip-directed 3D growth*, *the writing mode* that can also produce 3D structures in the repeating passages including walls, and *the stencil mode* that produces wall-like structures of various shapes. Manipulating all of them gives sufficient freedom to realize complex 3D designs. The phenomenological theory that we presented is robust and simple to help organize the 3D growth process to compete with the controlled 3D "drawing" usually provided by laser techniques in polymer based material.  General equations for arbitrary tip growth direction are quite simple to be able to solve the reverse problem of finding the appropriate stage motion while being given the desired morphology of the grown structure. Last but not least is that the vertical position of the mask is not important which greatly simplifies the movement protocol of the stage: 2D stage movement in its own plane might be enough for full 3D printing however complex.

**Acknowledgements**

This work was supported by the Global Frontier R&D Program of the Center for Multiscale Energy System (2011-0031561 and 2012M3A6A7054855) by the National Research Foundation (NRF) under the Ministry of Science, ICT and Future Planning, Korea

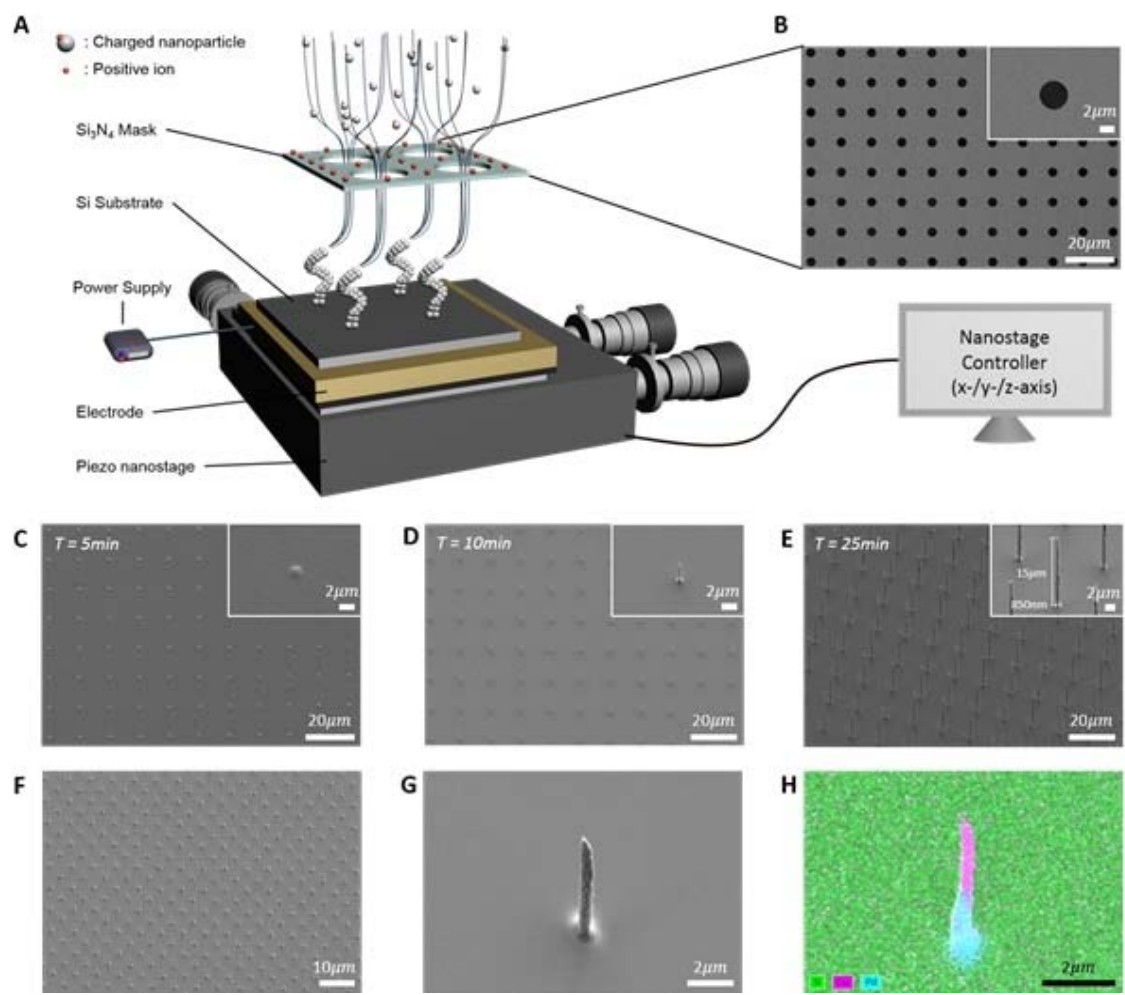

**Fig. 1. Assembling charged aerosols for 3D nanoprinting and nanocolumn structures.** (**A**) The aerosol based 3D nanoprinting method thorugh floating mask and piezoelectric nanostage. Negative potential applied substrate attracts positive ions and nanoparticles simultaneously injected from spark discharge generator. Ions with high mobility reach the substrate first and form distorted local electric field. After that, nanotarticles follow nanoscale focusing field lines and are printed on the substrate. (**B**) FE-SEM image of the floating mask. The diameter of mask holes is 4μm. (**C** to **E**) 3D nanocolumn structures with the flow of time. In the case of high aspect ratio nanocolumn (**E**), the stage translates vertically to z-axis (500nm min$^{-1}$) during the particle deposition after 15min staying. (**F**) Increasing of array density is possible by moving the stage horizontally. The stage moves in 7 steps and stays 10min for each step. (**G** and **H**) Different charged aerosols can be used in single fabrication process. (**G**) FE-SEM and (**H**) EDS image confirms successful fabrication.

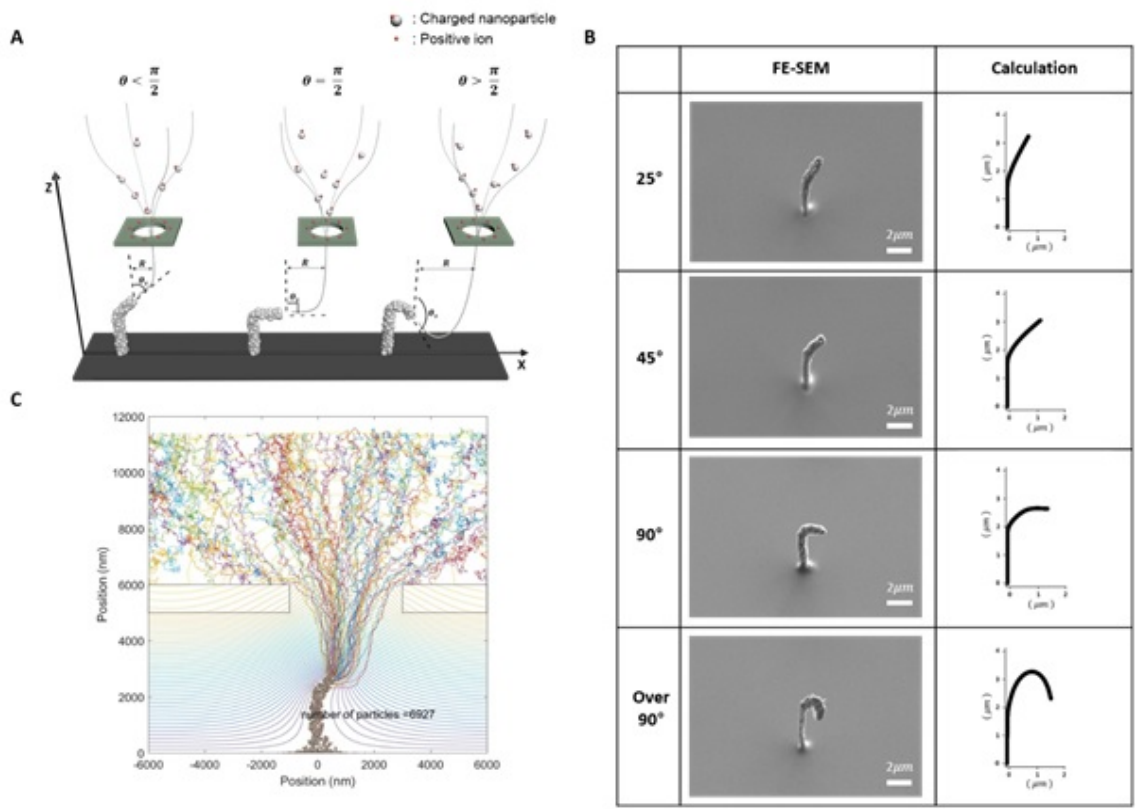

**Fig. 2. FE-SEM images and simulation result of 3D slanted structures having various slanted angles.** (**A**) Schematic of the growing process where nanoparticles come from the hole in the mask, following the guiding electric field line. (**B**) Experimental structures (left panel) against the ones, calculated by our theory (right panel), grown a different stage motion protocols that give different angles $\theta_s$ between the tip growth direction and the vertical direction. Lowest images show a counterintuitive downward growth direction predicted by our theory. (**C**) MATLAB numerical simulation of nanoparticle trajectories. The 3D slanted structure growth showing the nanoparticle trajectories in the focusing electric field. Ion accumulated surface charge density measured by KFM. (Fig. S4)

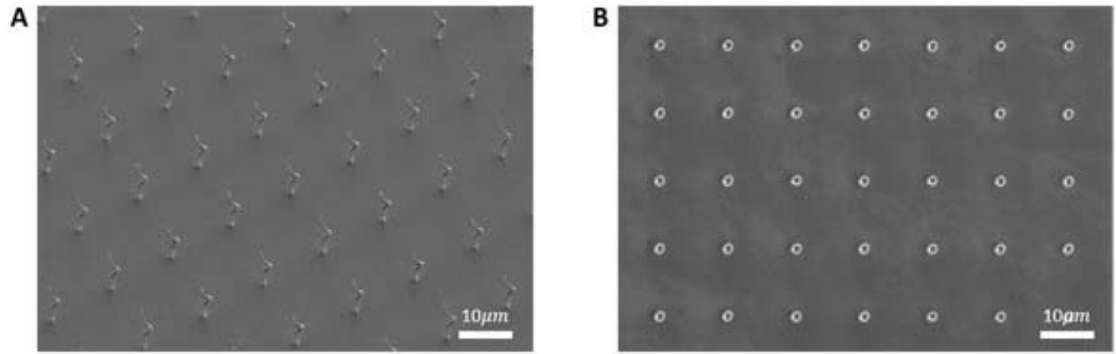

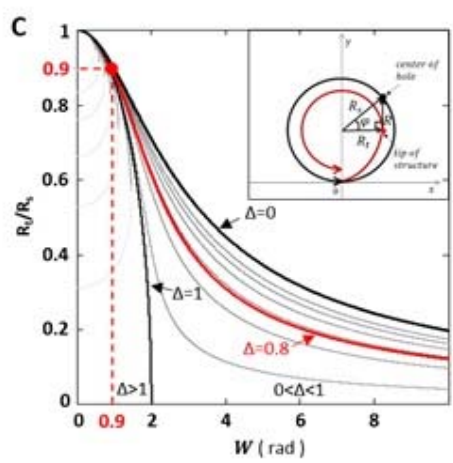
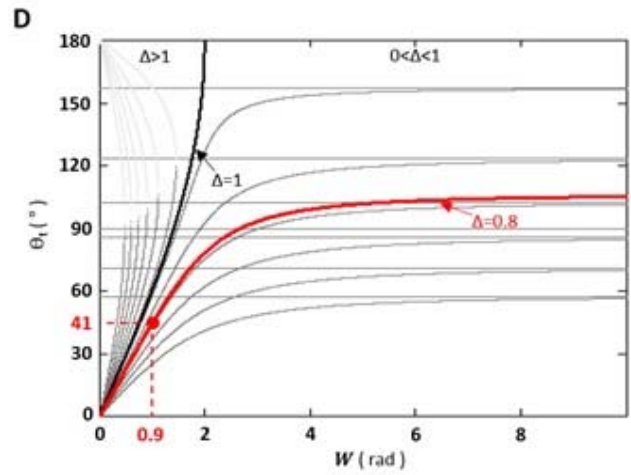

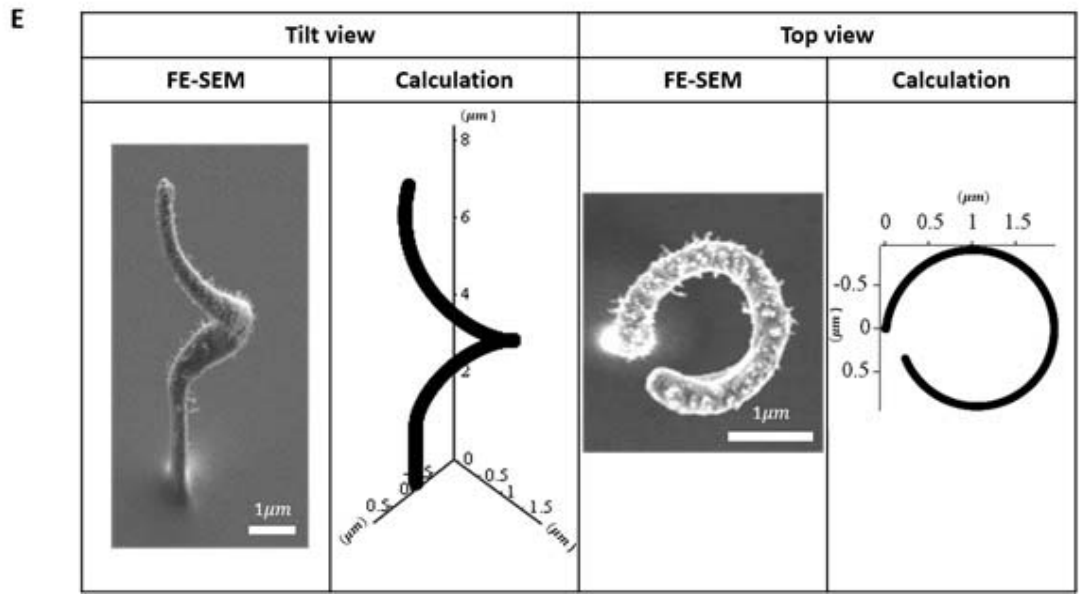
17

**Fig. 3**. **FE-SEM images and calculations from phenomenological theory of 3D helix structures.** (**A**) Helices 3D printed with a rotating stage in tilt view and (**B**) top view. (**C**) Normalized helix radius vs dimensionless angular velocity as a series of branches of the solutions of Eq. (3) for $\Delta < 1$ and $\Delta > 1$ (where $\Delta = R_s/L_p$ is the ratio of the stage rotation radius and the persistence length), which cases are separated by the thick black line. Inset gives the geometry of the steady state stage rotation and the relative tip position. The red line shows the parameter line in which the experimental helix parameters (numbers in red) lie (red dot); (**D**) helix pitch angle in degrees vs angular velocity. Horizontal lines present the saturation value $2\,\mathrm{asin}\,\Delta$ of the pitch angle at large angular velocity (recalculated into degrees) that follows from Eq. (3). The red line, red dot, and red numbers correspond to the experimental helix parameters. (**E**) Tilt and top view of theoretical (right panel) and experimental (left panel) helix structures presented *en mass* in (**A**) and (**B**).

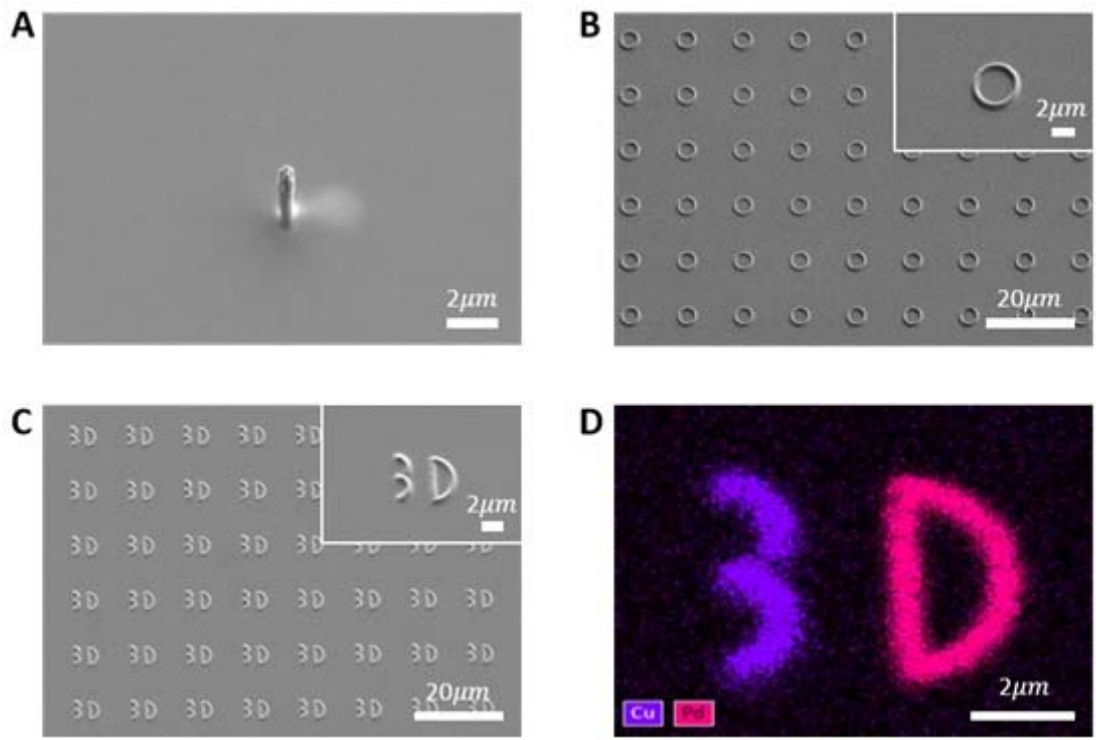

**Fig. 4. 3D structures created by 3D writing mode.** (**A**) Scattered nanoparticles represent that nanoparticles cannot follow existing nanocolumn when the stage translation velocity increases over certain value. The stage translate to x-axis (25nm s$^{-1)}$) after 10min deposition in first position. (**B**) Circular movement of the stage can manufacturing 3D cylinder structures of 4μm diameter. (**C- D**) 3D writing mode can write letters on the substrate through a programmed stage movement. (**C**) represents FE-SEM image of '3' and 'D' letters and EDS data (**D**) confirms they are written with different materials.

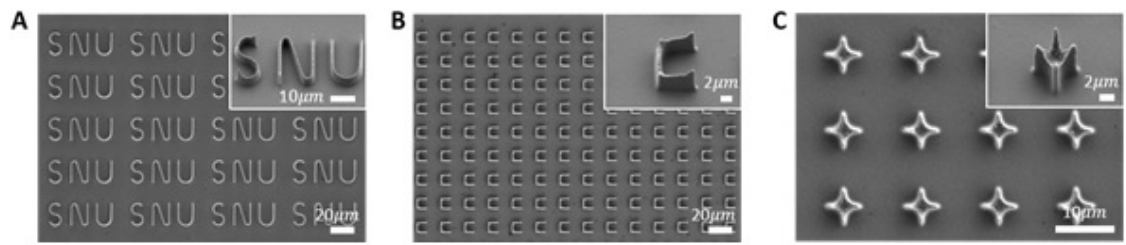

**Fig. 5. Various 3D structures depending on floating mask design.** (**A**) SNU (abbreviation of Seoul National University), (**B**) Korean alphabet, and (**C**) cross pattern.

**Supplementary Materials**

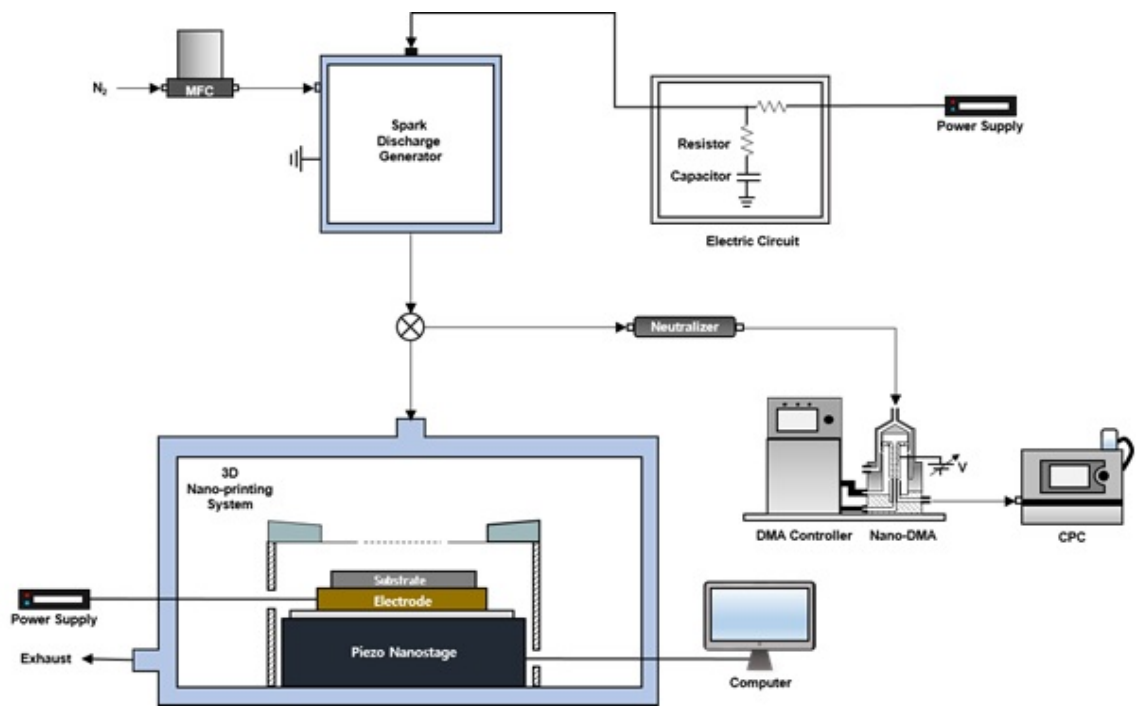

**Fig. S1. Experimental set-up for fabrication of multiscale 3D structures using an ion-induced focusing floating mask.** It is composed of a spark discharge generator for charged nanoparticles generation, SMPS, and 3D nano-printing system. The 3D nano-printing system consists of the floating mask, a substrate, an electrode, and a piezoelectric nanostage which is controlled by a computer.

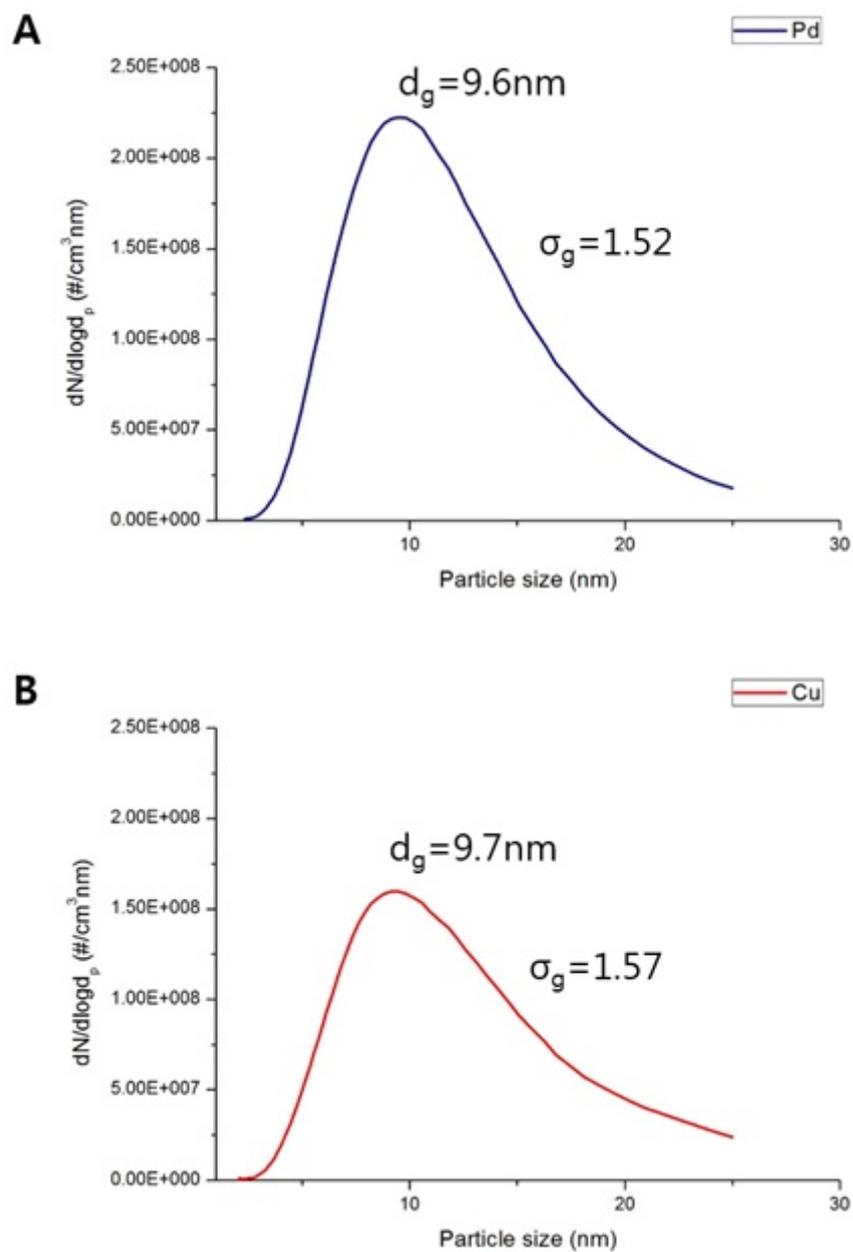

**Fig. S2. Size distribution of nanoparticles measured by SMPS.** (**A**) Pd and (**B**) Cu nanoparticles size distributions showing geometric mean diameter, geometric standard deviation, and total number concentration.

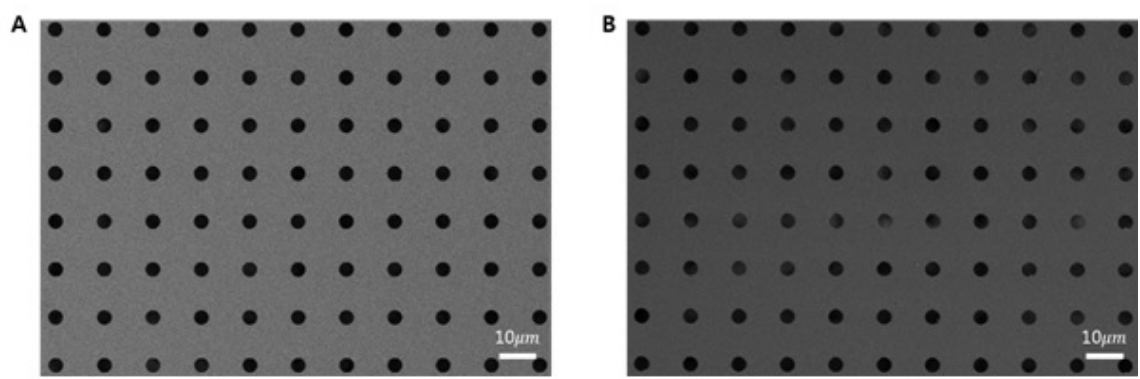

**Fig. S3. FE-SEM images of the floating mask.** (**A**) Before and (**B**) after use.

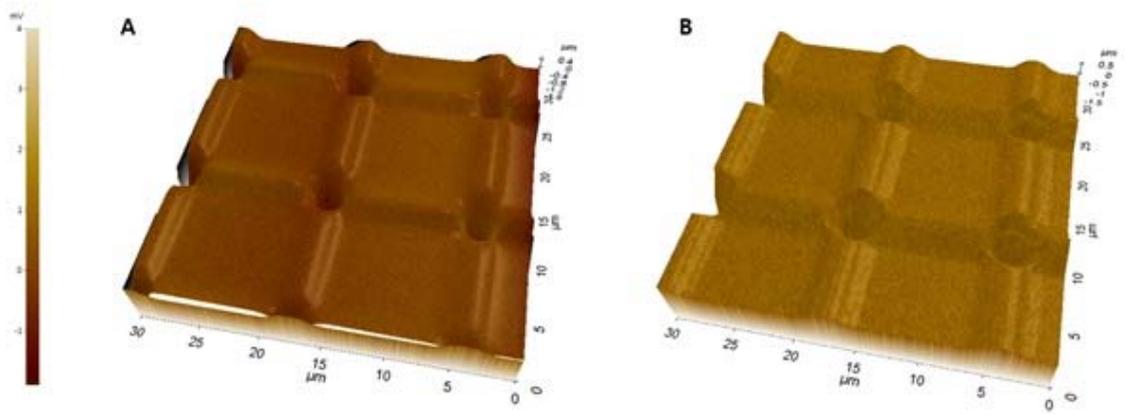

**Fig. S4. AFM topography images and KFM images of surface potential distribution on the floating mask.** (**A**) Before and (**B**) after use.

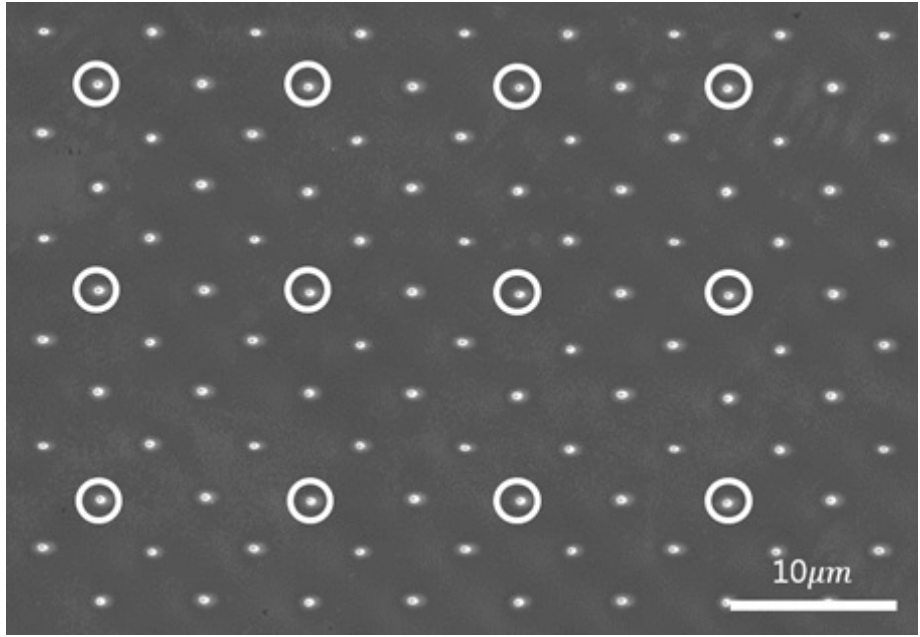

**Fig. S5. Top view of high array density created by moving the piezoelectric nanostage in 7 steps.** White circles are the first structures before moving the stage. Deposition time of each step is 10min.

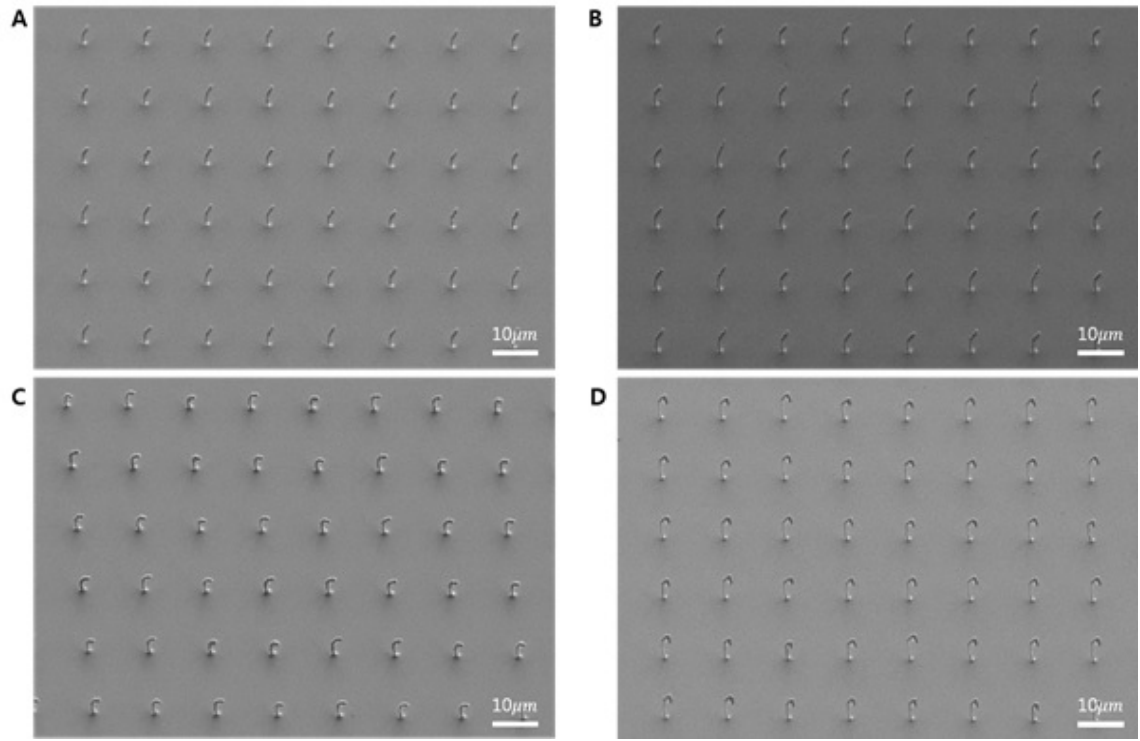

**Fig. S6. Tilt view with FE-SEM low magnification for each slanted angle of 3D structures.** (**A**) 25°, (**B**) 45°, (**C**) 90°, and (**D**) over 90°.

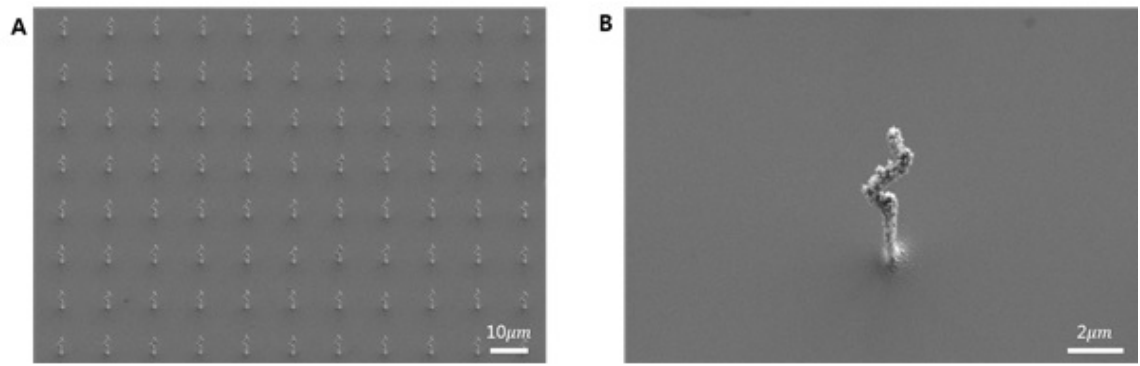

**Fig. S7. 3D zig-zag shape structures.** After 15min deposition in place, the piezoelectric nanostage jumped -1μm on the x-axis and stayed for 4min. Then it jumped to 1μm on the x-axis, stayed for 4min, and repeated one more time. After that, it jumped -1μm on the x-axis again and stayed for 4min. (**A**) Tilt view with low magnification and (**B**) one structure in the array.

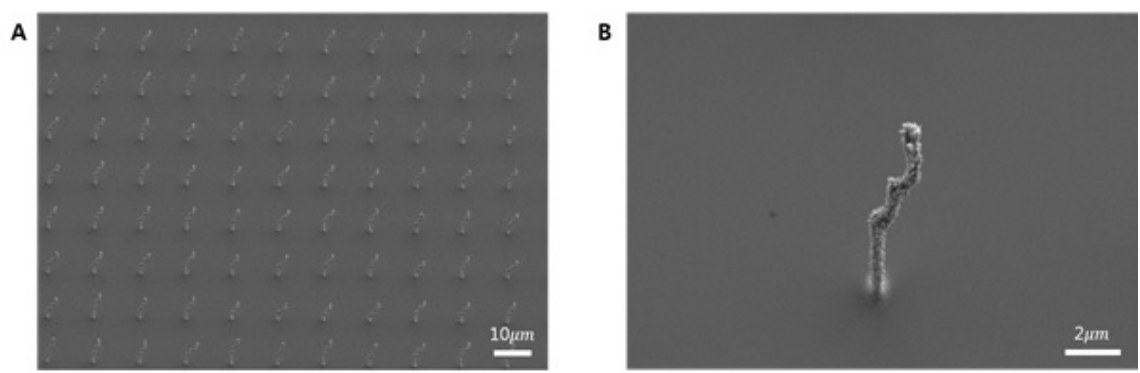

**Fig. S8. 3D stair shape structures.** After 20min deposition in place, the piezoelectric nanostage jumped 1μm on the x-axis, stayed for 7.5min and repeated this process one more time. (**A**) Tilt view with low magnification and (**B**) one structure in the array.

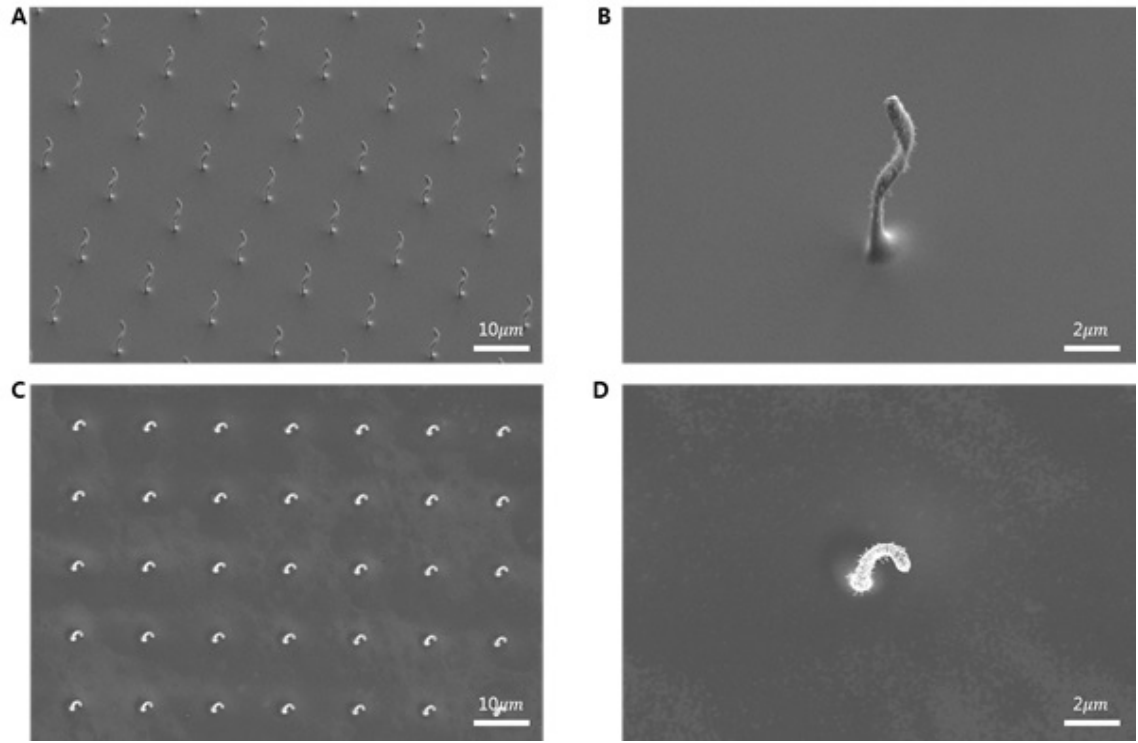

**Fig. S9. 180° 3D helix structures.** Piezoelectric nanostage rotation radius with 1μm, constant angular frequency with $4\pi$ h$^{-1}$, rotation angle with 180°, and persistence length with 1.12μm. (**A**), (**B**) Tilt view. (**C**), (**D**) Top view.

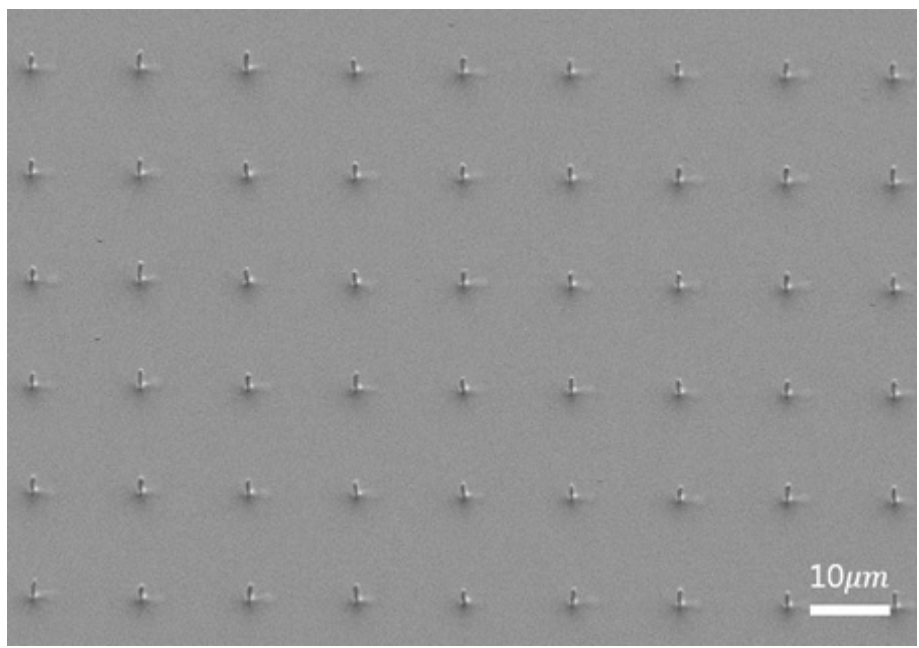

**Fig. S10. Tilt view with FE-SEM low magnification of 3D structures created by writing mode.** After 10min deposition, the charged nanoparticles are deposited on the surface of substrate instead of the tip of the structures because of movement speed of the piezoelectric nanostage, 25nm s$^{-1}$, on the x-axis.

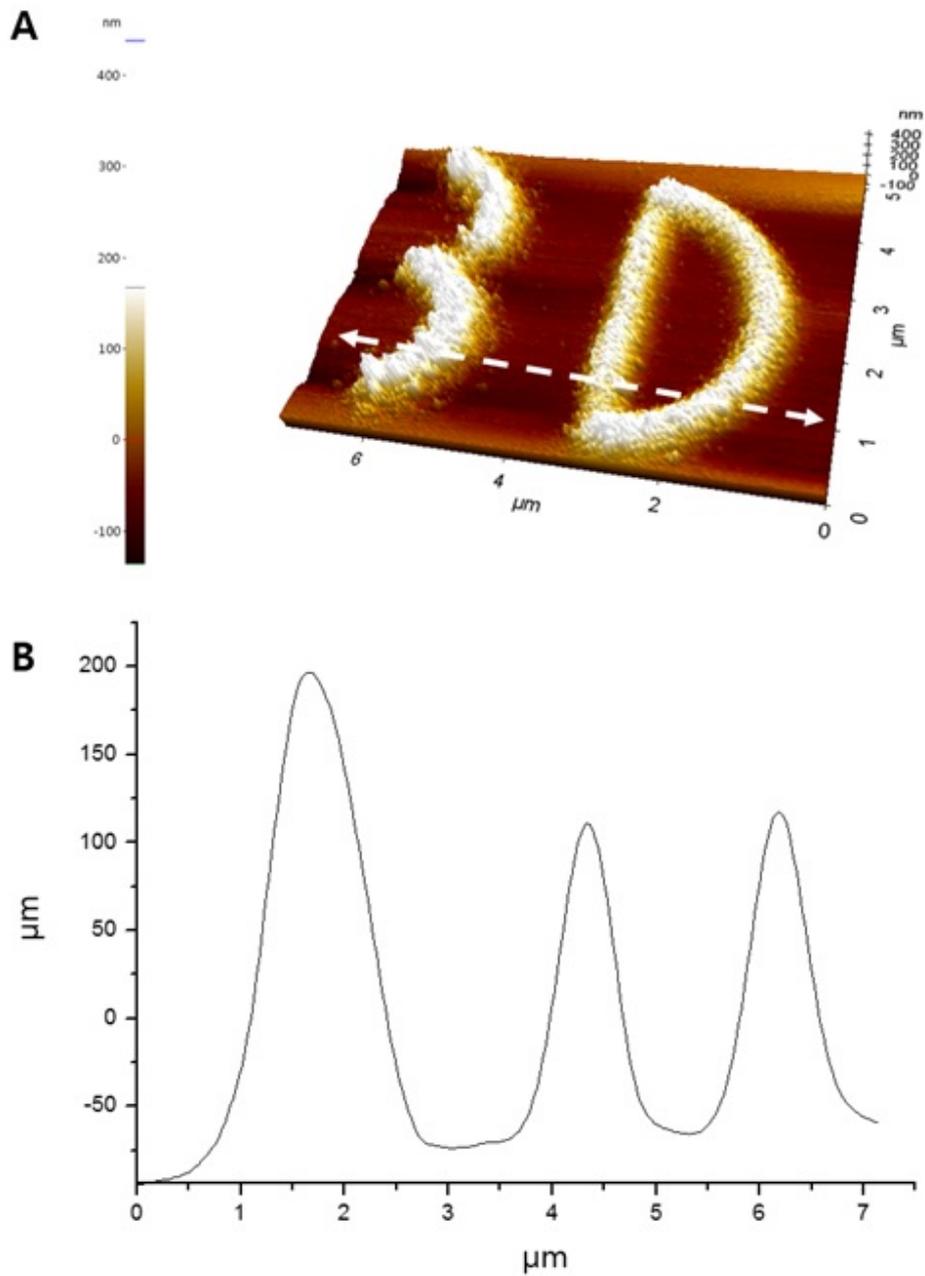

**Fig. S11. AFM topography image and height graph of '3D' structure in writing mode.** (**A**) AFM image of '3D' structure topography. (**B**) Height graph of '3D' structure following a white line at (**A**).